\documentclass[twocolumn,showpacs,preprintnumbers,amsmath,amssymb]{revtex4}

\usepackage{graphicx}
\usepackage{dcolumn}
\usepackage{bm}

\begin{document}             

\title{Collapse of the vortex-lattice inductance and shear modulus at the melting transition in untwinned $\rm 
YBa_2Cu_3O_7$.} 
\author{Peter Matl$^{1,*}$, N. P. Ong$^1$, R. Gagnon$^2$, and L. Taillefer$^{2,\dagger}$}      
\address{$^1$Department of Physics, Princeton University, Princeton, New Jersey 08544}
 
\address{$^2$Department of Physics, McGill University, 3600 University
Street, Montr\'eal, Quebec, Canada H3A 2T8}

\date{\today}      


\begin{abstract}
The complex resistivity $\hat{\rho}(\omega)$ of the vortex lattice in an untwinned crystal of 93-K $\rm 
YBa_2Cu_3O_7$ has been measured at frequencies $\omega/2\pi$ from 100 kHz to 20 MHz in a 2-Tesla 
field $\bf H\parallel c$, using a 4-probe RF transmission technique that enables continuous measurements 
versus $\omega$ and temperature $T$.  As $T$ is increased, the inductance ${\cal L}_s(\omega) ={\rm 
Im}\; \hat{\rho}(\omega)/ \omega$ increases steeply to a cusp at the melting temperature $T_m$, and then 
undergoes a steep collapse consistent with vanishing of the shear modulus $c_{66}$.  We discuss in detail 
the separation of the vortex-lattice inductance from the `volume' inductance, and other skin-depth effects.  
To analyze the spectra, we consider a weakly disordered lattice  with a low pin density.  Close fits are 
obtained to $\rho_1(\omega)$ over 2 decades in $\omega$.  Values of the pinning parameter $\kappa$ and 
shear modulus $c_{66}$ obtained show that $c_{66}$ collapses by over 4 decades at $T_m$, whereas 
$\kappa$ remains finite. 
\end{abstract}

\maketitle                   

\section{Introduction}\label{intro}
The investigation of the response of vortices in type II superconductors to an alternating driving current has 
had a long and fruitful history, beginning with the experiment of Gittleman and Rosenblum (GR) on PbIn and 
Nb alloys \cite{Gitt}.  In cuprate superconductors, these investigations have been performed on $\rm 
YBa_2Cu_3O_7$ (YBCO) over a broad range of frequencies extending from 100 kHz to a few THz.  At 
frequencies $\omega$ below a few 100 MHz, the vortex response is obtained by directly measuring the 
sample's complex resistivity $\hat{\rho}(\omega)$.  At microwave frequencies (10 to 100 GHz), cavity 
perturbation techniques have been the primary approach, although bolometric absorption techniques have 
been more useful at the high microwave end.  

The occurence of the vortex solid-to-liquid melting transition in the cuprates is one of the most interesting 
phenomena in the investigation of `vortex matter' in superconductors~\cite{Blatter}.  The melting transition at 
$T_m$ has been investigated in some detail by high-resolution measurements of the magnetization 
\cite{Zeldov,Welp}, heat capacity \cite{Schilling} and $dc$ resistivity \cite{Kwok,Safar}.  
In any solid-to-liquid transition, the key quantity of interest is the shear modulus which vanishes in the liquid.  
The resistivity profile ($\rho$ vs. $T$) in a crystal of $\rm Bi_2Sr_2CaCu_2O_8$ in which alternating 
strong and weak pinning channels are created by ion irradiation was interpreted as a loss of shear strength at 
$T_m$~\cite{Kes}.  The effect of the vanishing shear modulus at $T_m$ on $\rho$ when the field $\bf H$ 
is tilted relative to the twin boundaries in twinned YBCO crystals was also demonstrated 
~\cite{Kwokmod}.  

The standard approach to investigating how the shear modulus changes in a system undergoing a 
solid-to-liquid transition is by observing its {\em dynamical response} to an oscillating driving force.  The 
frequency dependence of $\hat{\rho}(\omega)$ in the vortex state in YBCO has been investigated by 
several groups at frequencies in the 1 to 100 MHz range.  Initial measurements using thin-film samples found 
that $\hat{\rho}(\omega) = \rho_1(\omega)+j\rho_2(\omega)$ displays strong, featureless dispersion at 
these frequencies \cite{Koch,Wu1,Anlage}.  However, the response in high-purity single crystals is 
qualitatively different.  Wu, Ong, Gagnon and Taillefer (WOGT) \cite{Wu2} investigated the $ac$ response 
in untwinned crystals of YBCO in the MHz range, and found that the inductance vs. $H$ exhibits a 
step-wise change at the melting field.  As a liquid cannot produce an inductive response, the step-wise 
change in the inductance is direct evidence for a collapse of the shear modulus $c_{66}$ at 
$T_m$.  

The 2-probe experiments of WOGT were performed at fixed $T$ versus $H$.  This approach cannot be 
extended to measurements in which $T$ is continuously varied in fixed $H$ because variations in the 
background signal (from $T$-dependent stray cable reactances) swamp the sample signal.  We report 
results obtained by the use of an RF 4-probe technique that circumvents this obstacle.  The 4-probe 
technique allows high-resolution impedance measurements versus each of the three quantities $\omega$, 
$H$ and $T$.  We find that the collapse of the inductance is even more abrupt when $T$ is scanned at fixed 
$H$.  While we largely confirm the {\em fixed}-$T$ results of WOGT, the swept-$T$ experiments provide 
a more direct picture of the behavior of the vortex resistivity and inductance as $T$ is increased above the 
melting line.  With the broader perspective, we correct a previous inference regarding a `remanent' shear 
modulus in the vortex liquid state.  A detailed discussion of how the vortex-lattice inductance is separated 
from the `volume' inductance is provided.  Together with the measurements of WOGT, the present results 
provide a complete experimental picture of the linear response of the pinned lattice in a high-purity cuprate 
crystal to an ac driving current in the MHz frequency range.  The two distinguishing features are the strong 
dispersion of $\hat{\rho}(\omega)$ observed even at low $\omega$ (just below the melting temperature 
$T_m$) and the striking abrupt collapse of the inductance at $T_m$.  

To understand the spectra, we have adopted a model in which the vortex lattice is weakly disordered by 
pinning to a low density of pins \cite{Schmid,Larkin}.  A mean-field solution proposed by Ong and Wu 
(OW)~\cite{Ong} is used to fit the measured resistivity spectra.  By measuring the damping viscosity 
$\eta(T)$ independently, we reduce the number of adjustable parameters to just 2 numbers at each $T$. 
We show that the OW solution can achieve close fits to $\rho_1(\omega)$ extending over 2 decades in 
$\omega$.  The fits enable us to find both $\kappa$ and $c_{66}$ (to within a multiplicative constant) at 
each $T$.  The latter exhibits a remarkable 4-decade collapse to zero within 1 K of $T_m$.   

In Sec. \ref{exper} a summary of the experimental approach is given. Section \ref{skin} discusses 
skin-depth effects and the `volume' inductance.  In Sec. \ref{fixedfreq}, we report measurements versus 
$T$ and $H$ at constant $\omega$.  The central results -- the spectra of $\rho_1$ and $\rho_2/\omega$ -- 
are reported in Sec. \ref{spectra}.  Section \ref{fits} explains the fits of the spectra to the weakly disordered 
lattice model, and Sec. \ref{modulus} discusses the temperature dependence of the shear modulus near 
$T_m$.  Appendix A describes the 4-probe technique, while Appendix B summarizes the impedance 
calculation for a sample with elliptical cross-section.  Appendix C summarizes the linear response of the 
weakly disordered lattice to an $ac$ current.

\section{\label{exper}Experimental details}
As in WOGT \cite{Wu2}, a weak, alternating current ${\bf J} {\rm e}^{j\omega t}$ is applied within the 
$\rm CuO_2$ plane $({\bf J} \parallel {\bf a})$ in the presence of a field $\bf H\parallel \hat{c}$.  The 
average velocity response of the vortices is detected as a complex resistivity $\hat{\rho}(\omega) = 
\hat{v}(\omega)B/J_0$.  

Because contact resistances on YBCO crystals are typically a fraction of an Ohm, they add a 
$T$-dependent background that dominates the sample signal.  In an experiment in which the field $H$ is 
swept at constant $T$ (as in WOGT), this large background may be subtracted, by reference to the 
zero-field data at each $T$.  However, when $T$ is swept at constant $H$, the contact background leads 
to substantial errors.  The problem is exacerbated by the $T$ dependence of the dielectric filling in the long 
coax cables.  We minimized this by using special cables with low-density Polytetrafluoroethylene (teflon) 
dielectric.  The higher propagation velocity (0.85$c$ versus 0.66$c$) is also an advantage.  

The 4-probe design was developed to remove the background altogether.  The sample (impedance 
$Z_s=R_s +j\omega L_s$) shunts the inner and outer conductors of both the incident and transmission 
coaxial cable (see Appendix A).  If $Z_s\ll Z_0$ (the line impedance), the incident signal is strongly 
reflected, and the transmitted signal is highly sensitive to small changes in $Z_s$.  By phase-detecting the 
latter we may determine $Z_s$.  By inserting broad-band transformers in both the incident and transmission 
cables, we can achieve true 4-probe measurement at frequencies up to 50 MHz (Appendix A describes the 
calibration procedure).

The measurements were made on a detwinned crystal of YBCO (of dimension $a$ = 1.11 mm, $b$ = 0.41 
mm, and $c$ = 55 $\mu$m), in which the $dc$ resistivity displays a sharp transition (width of 0.2 K) at 
$T_c$ = 93.3 K.  The RF resistivity was measured with the RF current density $\bf J\parallel a$, and $\bf 
H\parallel c$.  Although all measurements were performed in a screened room, radio broadcast signals are 
picked up as weak resonance lines in the spectra.  Parasitic line resonances, detected as weak resonances, 
are the hardest to eliminate.  However, they can be minimized by empirically adjusting the grounding strap 
configuration. 

With the 4-probe method, we may vary independently each of the variables $H$ (0-8 Tesla), $\omega$ 
(100 kHz to 20 MHz), and $T$.  To confirm that we are in the linear-response regime, we checked that 
curves of $\hat{\rho}(\omega)$ measured with incident power at -15.5 dBm and at 4.5 dBm are closely 
similar (except for a higher noise content in the -15.5 dBm curve for $\rho_2$).  We report all 
measurements in terms of the sample's complex resistivity $\hat{\rho}(\omega) = Z_s(\omega)(bc/a') \equiv 
\rho_1(\omega) + j\rho_2(\omega)$, where $a'$ = 0.68 mm is the separation of the voltage contacts.  In 
place of $\rho_2(\omega)$, it is preferable to discuss the (specific) inductance defined by ${\cal L}_s \equiv 
\rho_2/\omega$ (note that ${\cal L}_s =L_s(bc/a')$).  Dividing $\rho_2$ by $\omega$ isolates the 
divergent behavior of ${\cal L}_s$ as $\omega\rightarrow 0$, thereby highlighting the dramatic collapse of 
${\cal L}_s$ at the lattice melting temperature.  

\section{\label{skin}Skin-depth and volume inductance}

In zero magnetic field, the impedance $Z_s(\omega)$ of a superconductor (in the form of a cylinder) may be 
calculated from the 2D Helmholtz equation $(\nabla^2+\hat{\kappa}^2){\bf A}=0$, where 
$\hat{\kappa}^2 = \lambda^{-2}+ 2j\delta^{-2}$ (here, $\lambda$ is the London penetration length and 
$\delta$ the skin-depth determined by the quasiparticle conductivity).  

First, we consider the normal-state impedance ($T>T_c$) by letting $\lambda\rightarrow \infty$, and 
replacing $\delta$ by the normal-state skin-depth $\delta_n= \sqrt{2\rho_n/\omega\mu_0}$ (where 
$\rho_n$ is the normal-state resistivity and $\mu_0$ the vacuum permeability).  When $\omega$ is large 
enough that $\delta_n\ll \sqrt{\cal A}$, $J$ is confined to the skin-depth, and the effective $R_s(\omega)$ 
increases (${\cal A}$ is the cross-section area).  For a circular cross-section, $Z_s(\omega)$ is given by 
\cite{Marion}
\begin{equation}
Z_s(\omega) = \frac{\omega\mu_0\ell\delta_n}{2\pi r_0\sqrt{2j}}
\frac{J_0(\hat{x})}{J_1(\hat{x})},
\label{Zcirc}
\end{equation}
where $\hat{x} \equiv -jr_0\sqrt{2j}/\delta_n$, $J_m(\hat{x})$ is the Bessel function of order $m$, and 
$r_0$ ($\ell$) is the radius (length) of the sample. The case of elliptical cross-section is treated in Appendix 
B. 

Equation \ref{Zcirc} shows that, at large $\omega$, $R_s(\omega)\rightarrow \rho\ell/(2\pi r_0\delta_n)$, as 
expected. More important for our inductance discussion, the RF field energy at high $\omega$ is confined to 
the skin-depth.  Therefore, the inductance $L_s(\omega)$ decreases monotonically to zero as 
$\omega\rightarrow \infty$.  As this inductance expresses the RF energy stored in the volume of a {\em 
straight} cylinder, we refer to it as the `volume' inductance $L_{vol}(\omega)$ to distinguish it from the 
vortex-lattice inductance which is the main focus of this work (and from stray, geometric inductances that 
arise from coil-coil flux linkage).  From Eq. \ref{Zcirc}, we find that 
\begin{equation}
\lim_{\omega\rightarrow 0} L_{vol} = \frac{1}{2}\frac{\mu_0 \ell}{4\pi}.
\label{Ln}
\end{equation}
For a high-purity Cu wire ($r_0$ = 0.4 mm, $\ell$= 3 mm) at 300 K, $L_{vol}(\omega)$ equals 120 pH at 
low $\omega$, but falls to a tenth of this value at 20 MHz.  

Below $T_c$, the penetration of the RF fields (in zero $H$) is set by the London length, i.e. 
$\hat{\kappa}\rightarrow \lambda^{-1}$.  As the in-plane $\lambda$ in YBCO is very short ($\sim 0.2 
\mu$m), we may regard the stored RF field energy to be nominally zero.  Hence the change in $L_s$ (in 
zero field) between $T_c$ and, say 60 K, provides a direct measurement of $L_{vol}$.  Applying Eq. 
\ref{Ln} to a cylinder with circular cross-section area equal to that in our crystal, we find that 
$\lim_{\omega\rightarrow 0} L_{vol}$ = 34 pH (corresponding to ${\cal L}_{vol} = 1.2\times 10^{-13}$ 
Hcm).  This tiny inductance may be measured with about 2$\%$ resolution by our 4-probe 
method.  

In the resistive state ($H\gg H_{c1}$), the RF penetration is determined by $\rho_1(\omega)$ (i.e. 
$\hat{\kappa}^2 = 2j\delta_f^{-2}$, where the flux-flow skin-depth $\delta_f= 
\sqrt{2\rho_1/\omega\mu_0}$).  As $\rho_1$ may be as large as 8 $\mu\Omega$cm for $H$= 2 T, the 
volume inductance $L_{vol}$ nearly recovers its full value given by Eq. \ref{Zcirc}.  However, in our 
measurements, the inductance increase is actually far in excess of $L_{vol}$.  The excess inductance arises 
purely from the vortex lattice.  

\begin{figure}
\includegraphics[width=4cm]{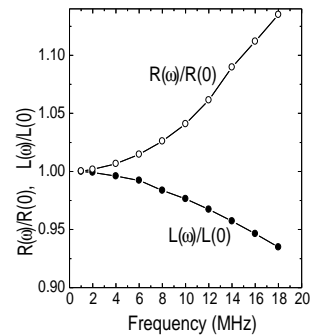}
\caption{\label{elliptical} Calculated impedance $Z_s(\omega)= R_s(\omega)+ j\omega L_s(\omega)$ of a 
conducting cylinder of elliptical cross-section and resistivity $\rho = 4{\rm \mu\Omega cm}$.  The 
inductance $L_s$ is entirely volume inductance $L_{vol}$. The parameters $\xi_0\simeq 0.108$ and $h= 
250 \mu$m are selected to match the sample cross-section.}
\end{figure}

Returning to the resistive component $R_s(\omega)$, we have converted it to $\rho_1$ by the assumption 
that $\bf J$ is uniform over the cross-section (i.e. $\rho_1(\omega)= R_s(\omega)bc/a'$).  However, if 
$\delta$ is comparable to the half-thickness $c/2$, $R_s(\omega)$ is significantly enhanced by the 
skin-depth effect, and $\rho_1$ will be overestimated.  We now address this concern.  As the aspect ratio 
$b/c\simeq 8.3$, we cannot rely on Eq. \ref{Zcirc} for quantitative guidance.  In Appendix B, we have 
calculated $Z_s(\omega)$ for a normal-state sample with an elliptical cross-section using Mathieu functions.  
The variation of $R_s(\omega)$ and ${\cal L}_{vol}(\omega)$ is shown in Fig. \ref{elliptical}) for a sample 
with this high ellipticity.  We assumed the normal-state resistivity $\rho_n = 4 
\mu\Omega$cm. 

As the results depend only on $\delta_n$, values for $R_s(\omega)$ and $L_{vol}(\omega)$ at other values 
of $\rho_n$ are obtained by re-scaling the $x$-axis.  For example, at 85.9 K, $\rho_1$ measured in a field 
of 2 T with $\omega/2\pi = 3$ MHz equals $\sim 1 \mu\Omega$cm.  Re-scaling the frequency axis in Fig. 
\ref{elliptical} by a factor of 4, we see that at $\omega/2\pi$ = 12 MHz, $R_s(\omega)$ and 
$L_{vol}(\omega)$ deviate by only 5$\%$ and 3$\%$, respectively, from their low-frequency (i.e. 
uniform-distribution) values.  At the lowest $T$ ($<$ 84 K), however, the deviations may be up to 20$\%$ 
for $\omega/2\pi>$10 MHz, but these low $T$-high-$\omega$ data will not play any significant role in our 
analysis.  Throughout the low-$\omega$ region near the melting transition, the correction is less than 0.5 
$\%$, so that the uniform-$J$ assumption is valid.  

\section{\label{fixedfreq} Fixed-frequency measurements}
\begin{figure}
\includegraphics[width=7cm]{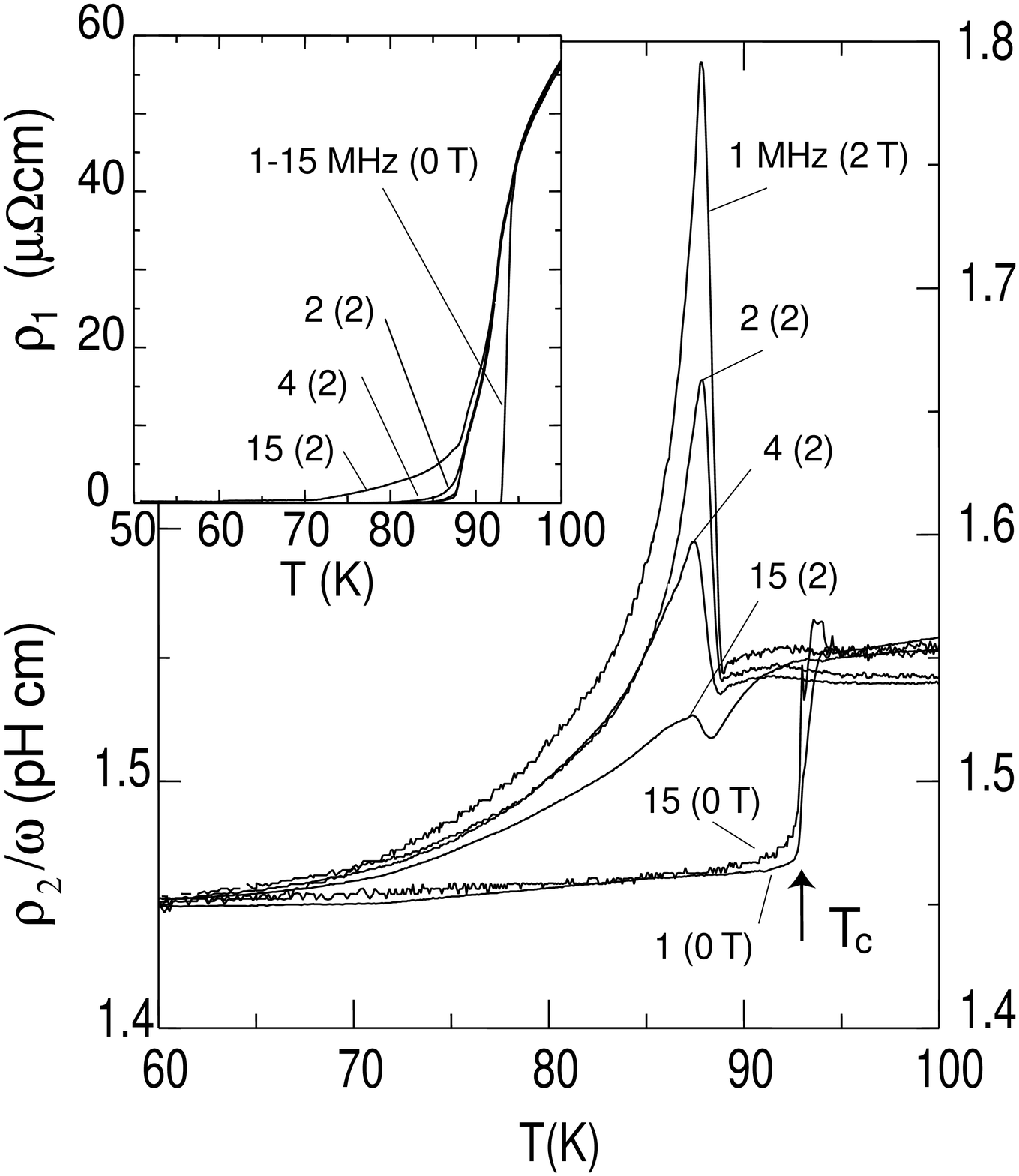}
\caption{\label{LvsT} (Main panel) Temperature dependence of the total sample inductance ${\cal 
L}_s(\omega,T,H)= \rho_2(\omega,T,H)/\omega$ measured in a 2-Tesla field ($\bf H\parallel c$) at the 
frequencies indicated.  Here, `15' means 15.4 MHz and $H$ (in Tesla) is in parantheses. As $T$ 
approaches $T_m$ from below, ${\cal L}_s$ (measured at 1-4 MHz) rises to a sharp cusp before 
collapsing to the value ${\cal L}_{vol}$ observed in the normal state.  For comparison, the zero-field traces 
taken at 1 and 15.4 MHz are also shown.  The inductance scale has an arbitrary origin (because of the term 
$L_{bg}$ in Eq. \ref{Lobs}).  The portion  of the curve 1 (0T) below 90 K serves as our reference line.  
Inductance signals measured relative to this reference are identified as contributions from the sample.  The 
step-change at $T_c$ in zero $H$ reflects the Meissner expulsion of the stored RF field energy (reduction 
of volume inductance ${\cal L}_{vol}$).  The inset shows $\rho_1$ measured at the same 
$\omega$.}
\end{figure}

All the data reported here are taken either in zero field or in a fixed field of 2 T. In the resistive state, the 
inductance of YBCO varies strongly with $\omega$ and $T$.  We first display the variation of ${\cal 
L}_s(\omega,T,H)$ versus $T$ measured at selected $\omega$ in a 2-Tesla field as well as in zero field 
(main panel of Fig. \ref{LvsT}).  The inset shows the corresponding curves for $\rho_1$.  

In the main panel, we have plotted the specific inductance ${\cal L}_s = \rho_2/\omega$ with an arbitrary 
origin, because the total observed inductance $L_{obs}$ is the sum of the sample contributions ($L_{vol}$ 
and the vortex lattice term $L_v$) and an arbitrary background $L_{bg}$ that arises from small 
uncertainties in setting `null' in the phase setting of the lock-in amplifiers), viz.
\begin{equation}
L_{obs} = L_v + L_{vol} + L_{bg}
\label{Lobs}
\end{equation}
We take the portion of the curve at 1 MHz-0T below 90 K as our reference, and identify sample inductance 
signals relative to this reference curve. 

\begin{figure}
\includegraphics[width=7cm]{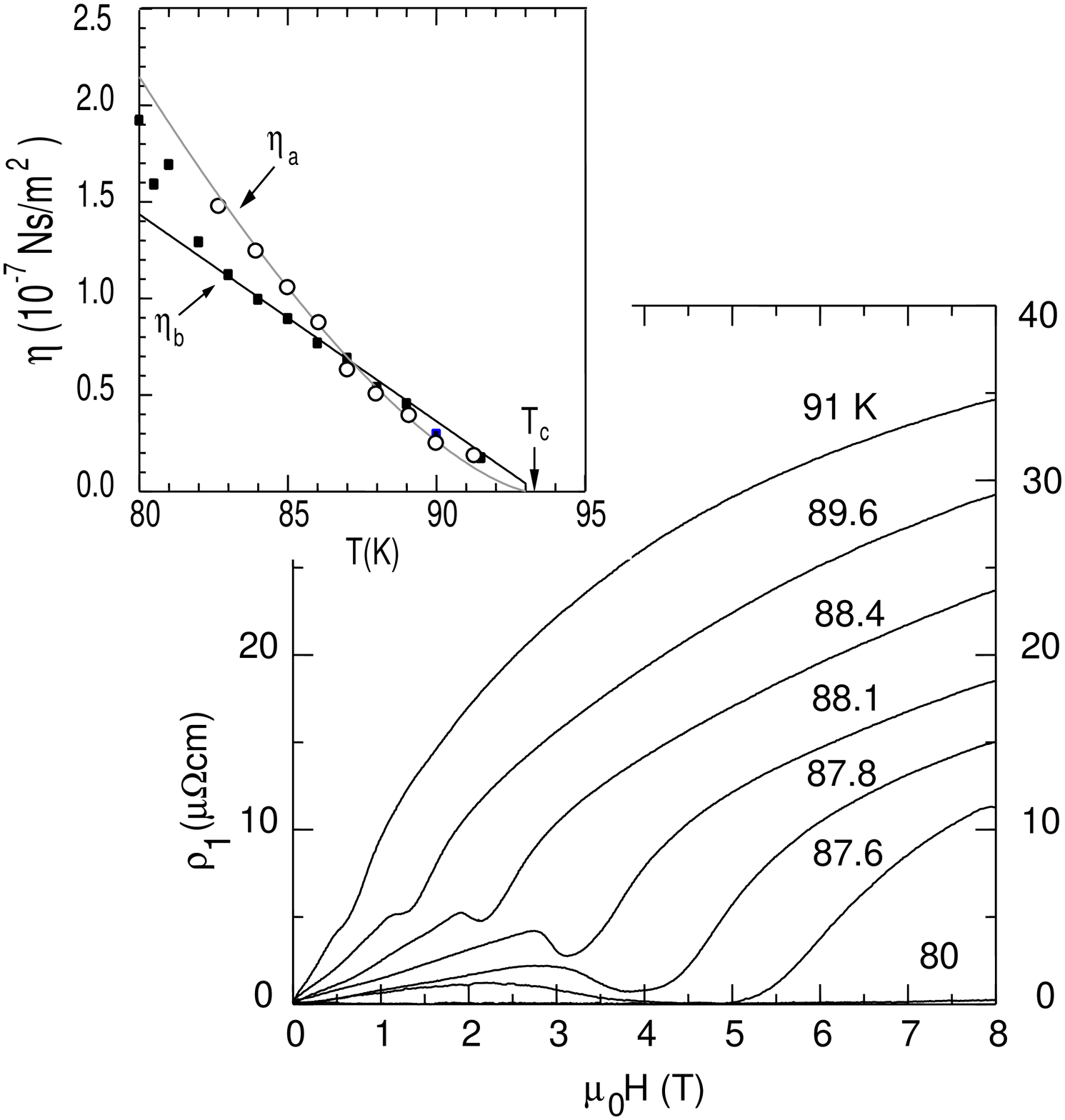}
\caption{\label{rhoeta} The high frequency resistivity $\rho_1(\omega)$ measured at $\omega$ = 15.4 
MHz versus applied field $\bf H\parallel \hat{c}$ at temperatures close to $T_c$ in untwinned YBCO ($\bf 
J\parallel \hat{c}$).  The minimum is caused by the `peak effect' which appears close to the melting field 
$H_m$.  At each $T$, the initial slope of $\rho_1$ vs. $H$ is used to determine the viscosity $\eta_a(T)$.   
The $T$ dependence of $\eta_a$ is plotted in the insert ($\eta_b$ obtained from a second sample is also 
plotted).  Near $T_c$, $\eta_a\sim(1-T/T_c)^{1.5}$, while $\eta_b\sim (1-T/T_c)$. 
}       
\end{figure}
We first discuss the 1 MHz curve in zero field. As $T$ increases from 60 K, ${\cal L}_s(\omega_1,T,0)$ is 
nearly $T$-independent until, at $T_c$, it undergoes a step-increase to the normal-state value 
($\omega_1/2\pi$ = 1 MHz).  Clearly, the step corresponds to the collapse of the Meissner effect at $T_c$, 
as discussed above.  The difference between ${\cal L}_s(\omega_1,T,0)$ measured at 100 and 60 K gives 
${\cal L}_{vol} = 1.0\;\times 10^{-13}$ Hcm, rather close to the value 1.2 $\times 10^{-13}$ Hcm 
estimated below Eq. \ref{Ln}.  Hence we identify the step change in the curve ${\cal L}_s(\omega_1,T,0)$ 
with the maximum (specific) volume inductance ${\cal L}_{vol}$ of our sample. The curve ${\cal 
L}_s(\omega_{15},T,0)$ measured at $\omega_{15}/2\pi$ = 15.4 MHz shows a closely similar step, as 
expected from Meissner flux-expulsion. 

\begin{figure}
\includegraphics[width=6cm]{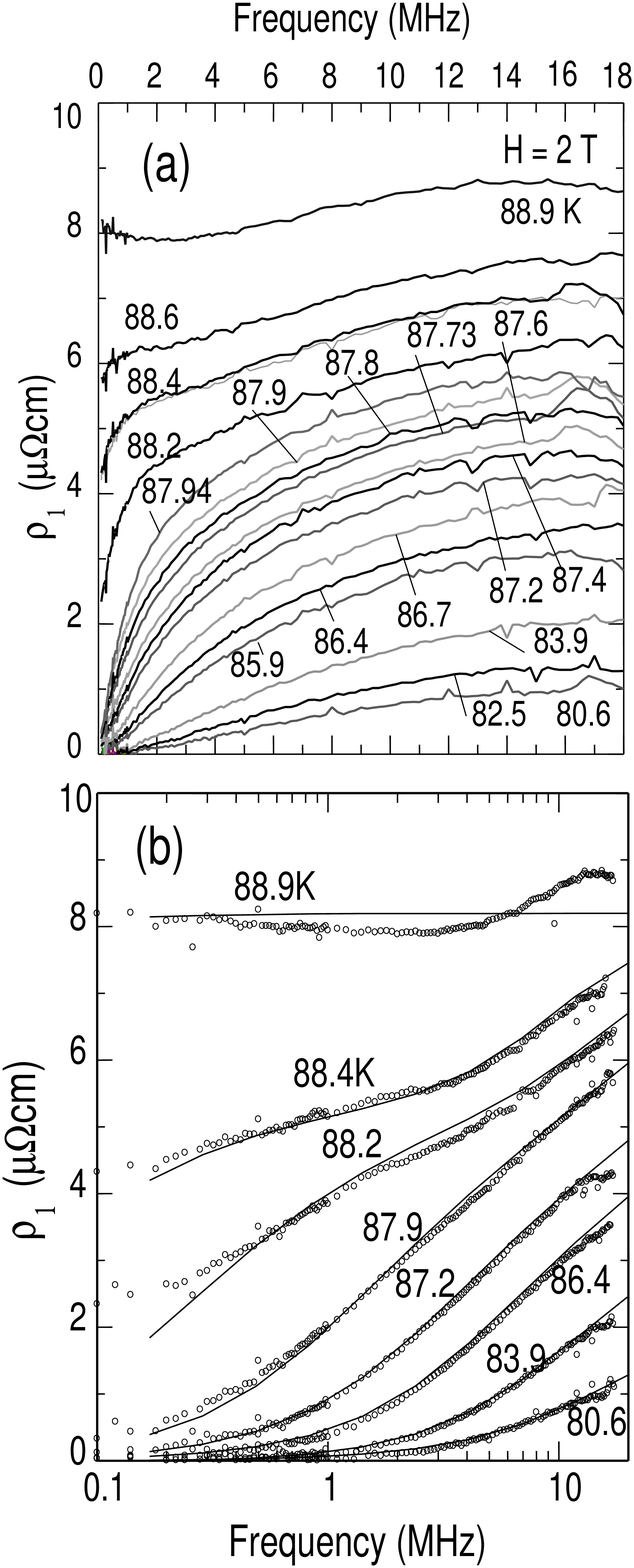}
\caption{\label{rho1} (a) The frequency dependence of $\rho_1(\omega,T,H_0)$ at temperatures 80.6 to 
88.9 K in a 2-Tesla field ${\bf H}_0\parallel{\bf c}$.  The curves for $\rho_1$ become increasingly 
$\omega$ dependent as $T_m (\sim$ 88.5 K) is approached from below.  In the vortex liquid state above 
$T_m$, $\rho_1$ rapidly approaches a nominally $\omega$-independent value (curve at 88.9 K).  (b) A 
subset of the data in upper panel plotted against $\log \omega$ to show the low-$\omega$ region.  The solid 
lines are fits discussed in Sec. \ref{fits}.
}       
\end{figure}

When a field $H_0$= 2 T is applied, the sample inductance is dominated by that of the vortex lattice.  We 
follow the curve at 1 MHz and 2 T, starting at 60 K.  As $T$ increases, ${\cal L}_s(\omega_1,T,H_0)$ 
increases rapidly, rising to a cusp just below the melting temperature of the lattice $T_m\simeq$ 88.5 K.  
Curves at higher frequencies show a similar but reduced peak.  Above $T_m$, ${\cal L}_s$ collapses 
rapidly to a value very close to the normal state value.  Because of the finite value of the RF vortex resistivity 
$\rho_1$ (inset), expulsion of the RF fields from the interior is now determined by the flux-flow skin-depth 
$\delta_f= \sqrt{2\rho_1/\mu_0\omega}$ ($\delta_f\gg \lambda$).  At these temperatures, $\delta_f > c/2$, 
so the RF fields are uniform in the vortex liquid state; there is no change in ${\cal L}_s$ in crossing $T_c$ = 
93.5 K.  

The sharp increase in ${\cal L}_s$ as $T$ approaches $T_m$ from below is one of our main results.  The 
collapse of the inductance above $T_m$ is the constant-$H$-scan version of the collapse observed by 
WOGT in constant-$T$ scans versus $H$.  

[As mentioned in Sec. \ref{intro}, the present swept-$T$ scans provide a much clearer picture of how the 
inductance evolves into the normal-state value.  In WOGT \cite{Wu1}, the finite value of ${\cal L}_s$ 
above the melting field $H_m$ was interpreted as evidence for a small, residual shear rigidity in the vortex 
liquid state (their Fig. 1b).  This inference is incorrect.  As discussed here, the residual inductance above 
$H_m$ is simply the volume inductance ${\cal L}_{vol}$.  Within our resolution, the shear rigidity in the 
vortex liquid state is zero.  This removes the need for the arbitrary `background' liquid term $\rho_2^b$ 
introduced by WOTG \cite{Wu2}.]

At our highest frequencies, the lattice response is predominantly resistive.  As discussed below, the friction 
term in this limit dominates the lattice forces and pinning forces, so that $\rho_1$ is proportional to the 
reciprocal of the viscosity $\eta(T)$.  Figure \ref{rhoeta} (main panel) displays the field dependence of 
$\rho_1(\omega)$ measured at the frequency $\omega/2\pi$ = 15.4 MHz at temperatures close to $T_c$ 
(the current is along the $a$-axis).  At each $T$, we have fitted the initial slope to the (free) flux-flow 
expression $\rho_f = B\phi_0/\eta$ in order to find $\eta(T)$ (plotted in the insert)  [$\phi_0 = h/2e$ is the 
flux quantum].   We measured $\eta(T)$ in 2 untwinned crystals.  In the geometry ${\bf J\parallel a}$, the 
uncertainty in determining $\eta_a$ is large because the broad `peak effect' restricts the range of $H$ over 
which $\rho_1$ vs. $H$ is truly linear.  More reliable results are obtained (for $\eta_b$) in the geometry 
${\bf J\parallel b}$ where the peak effect is much narrower.  The two sets of viscosity $\eta_a$ and 
$\eta_b$ are shown in Fig. \ref{rhoeta} (insert).  The viscosity data are used in the fits discussed in Sec. 
\ref{fits}.

\section{\label{spectra} Frequency dependence of the complex resistivity}
Figure \ref{rho1} displays the frequency dependence of $\rho_1(\omega,T,H_0)$ at temperatures from 
80.6 to 88.9 K ($H$ is fixed at 2 T).  At temperatures $T<T_m\sim$ 88.5 K, the value of $\rho_1$ in the 
limit $\omega\rightarrow 0$ is zero.  However, above $T_m$, the limiting value becomes finite.  At the 
highest $\omega$, $\rho_1$ approaches the Bardeen-Stephen value for free flux-flow $\rho_f$ (inset of Fig. 
\ref{LvsT}).

Below $T_m$, $\rho_1$ is strongly dispersive.  Close to $T_m$, $\rho_1$ displays a crossover from a 
sharply increasing region at low $\omega$ to a gradual region at high $\omega$.  The `knee' feature 
separating the two regions rapidly moves to high frequencies as $T$ falls below $T_m$ by a few K, 
eventually moving out of our frequency window. 

\begin{figure}
\includegraphics[width=9cm]{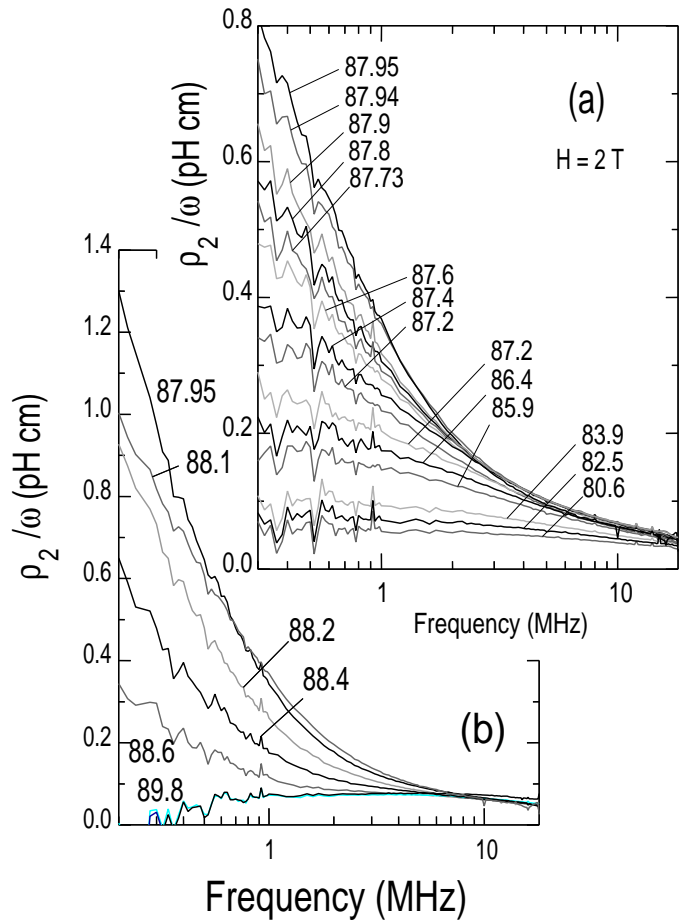}
\caption{\label{Llogf} The frequency dependence of the sample inductance ${\cal L}_s(\omega,T,H_0)$ 
measured in a 2-T field ${\bf H}_0\parallel{\bf c}$.  ${\cal L}_s$ (referenced to the zero-field inductance 
curve) is the sum of ${\cal L}_v$ and ${\cal L}_{vol}$.  The upper and lower panels are measurements 
taken below and above 87.95 K, respectively.  As $T\rightarrow T_m$ from below (a), the vortex term 
${\cal L}_v$ increases sharply, especially at low $\omega$.  Just below $T_m$ (b), it collapses to zero, 
leaving the nominally flat spectrum that we identify as the volume inductance ${\cal 
L}_{vol}$.}
\end{figure}

Figure \ref{Llogf} displays ${\cal L}_s(\omega,T,H_0)$ versus $\omega$ for temperatures below and 
above $T_m$ (with $H_0$ = 2 T).  As explained following Eq. \ref{Lobs}, we identify the sample 
inductance using the zero-field curve (below 90 K) as reference, viz. 
$${\cal L}_s(\omega,T,H) = [L_{obs}(\omega,T,H)-L_{obs}(\omega,T,0)](bc/a').$$  As discussed in 
Sec. \ref{fixedfreq}, ${\cal L}_s$ rises to a sharp cusp as $T\rightarrow T_m^-$, before undergoing a 
steep collapse above $T_m$.  The most pronounced dispersion occurs at low $\omega$ where ${\cal 
L}_s$ shows divergent behavior in the limit $\omega\rightarrow 0$.  

Although the spectra of the inductance are not used in our fitting process in Sec. \ref{fits}, we discuss 
various contributions to the observed inductance signal (which includes the volume inductance) for the sake 
of completeness.  We may estimate ${\cal L}_{vol}(\omega,T,H)$ by appealing to the observed $\rho_1$ 
and the calculated curves in Fig. \ref{elliptical}.  Note that ${\cal L}_{vol}$ can never exceed 0.12 pHcm.  
At high frequencies, $L_v$ is negligible, so that the curve at 15.4 MHz (2T) in Fig. \ref{LvsT} may be taken 
as that for ${\cal L}_{vol}$ vs. $T$ at large $\omega$ (we ignore the slight dip near $T_m$).  Using the 
values of $\rho_1$ measured at 15.4 MHz (Fig. \ref{rho1}), we find this interpretation is consistent with the 
calculations in Fig. \ref{elliptical}.

We may also use the observed values of $\rho_1$ to estimate ${\cal L}_{vol}$ at low frequencies.  At the 
temperatures of interest (85-90 K), $\rho_1$ is sufficiently large, even at 200 KHz, to satisfy the 
uniform-$J$ condition, i.e. ${\cal L}_{vol}$ is close to its maximum value 0.12 pHcm.  With increasing 
$\omega$, ${\cal L}_{vol}$ decreases slowly to its high-$\omega$ value as seen in Fig. \ref{LvsT}.  For 
example, at 88.2 K (in Fig. \ref{Llogf}b), our estimate of the volume inductance spectrum is a curve that 
starts at 0.12 pHcm at low $\omega$ and decreases slowly to $\sim$0.06 pHcm at high $\omega$. The 
divergent vortex term is easy to distinguish from this small, featureless background.   

This concludes the purely experimental part of the report.  The spectra in Figs. \ref{rho1} and \ref{Llogf}, 
complemented by the measurements of WOTG~\cite{Wu2}, constitute a rather complete quantitative 
description of the linear-response of the vortex lattice in untwinned YBCO at RF frequencies.  In the 
following sections, we interpret the measurements within a specific model, in order to extract the temperature 
dependence of the vortex-lattice shear modulus.

\section{\label{fits} Fits to the spectra}
\subsection{Discussion of the model}
In the model used by GR \cite{Gitt}, the vortex equation of motion is
\begin{equation}
\eta \dot{\bf u} + \kappa{\bf u} = {\bf J\times \hat{z}} \phi_0,
\label{GR}
\end{equation}
where $\bf u$ is the vortex displacement from equilibrium, and $\eta$ and $\kappa$ are the damping 
viscosity and the Labusch parameter, respectively, and $\bf B\parallel \hat{z}$.  Measurements of 
$\hat\rho$ in thin-film samples \cite{Koch,Wu1} and in untwinned crystals \cite{Wu2} of YBCO are in 
strong disagreement with the Lorentzian response $\hat{\rho} = (B\phi_0/\eta)[1-j\omega_p/\omega]^{-1}$ 
predicted by Eq. \ref{GR} (here $\omega_p\equiv \kappa/\eta$).  In particular, the dramatic dispersion 
observed when the lattice melting line is approached, in either constant-$H$ scans (WOGT) or in the 
constant-$T$ scans reported here, cannot be reproduced by Eq. \ref{GR}, even if one replaces $\kappa$ 
by an `effective' parameter that depends on $\omega$ and $B$.

In the context of the vortex glass~\cite{Fisher}, Dorsey \cite{Dorsey} has derived scaling relations for the 
$\omega$-dependent conductivity $\hat{\sigma}(\omega)$ at the vortex glass-to-liquid transition, viz. 
$\hat{\sigma}(\omega)\sim (-i\omega)^{\alpha}$.  In {\em thin-film} samples of YBCO, the measured 
spectra do in fact display power laws in $\omega$ over 2-3 decades (1-500 MHz)~\cite{Wu1}.  The 
situation in untwinned crystals is very different, however.  The spectra reported here (and in 
WOGT~\cite{Wu2}) do not follow power laws at all.  Further, the melting transition is abrupt and weakly 
first-order rather than the continuous transition of the vortex glass.  Under the high-purity conditions found in 
untwinned YBCO crystals, the observed spectra  {\em are more consistent with that of a weakly disordered 
vortex lattice with random pins} than a vortex glass (as we now show).

In their theory of the depinned vortex lattice, Schmid and Hauger \cite{Schmid} and Larkin and Ovchinikov 
~\cite{Larkin} represented the rigidity of the moving periodic structure by the lattice force matrix ${\bf 
D}^s_{l,m}$.  As discussed below, the non-Debye spectrum of $\hat{\rho}(\omega)$ and its rapid change 
vs. $T$ in the vicinity of $T_m$ reflect a rapidly changing length-scale in the problem.  To describe this, it is 
necessary to retain the $\bf q$ (wavevector) dependence of the Fourier transform $D_s({\bf q})$ of ${\bf 
D}^s_{l,m}$. (Clem and Coffey \cite{Clem} have shown that approximating $D_s({\bf q})$ by a constant 
$D$ merely reproduces a Debye-like spectrum.)  With inclusion of the pinning forces, the equation of 
motion is~\cite{Schmid}
\begin{equation}
\eta{\dot{\bf u}}_{\bf l} + 
\sum_{{\bf l}'} \bf D_{{\bf l,l}'}\cdot \bf u_{{\bf l}'}
+ \kappa\sum_{{\bf i}} {\bf u_i} \delta_{{\bf l,i}} 
= {\bf J\times \hat{z}} \phi_0,
\label{eq:eta}
\end{equation}
where the sites with pins are indexed by $\bf i$ (the pinning force is short-ranged).  To match the strong 
dispersion observed, OW assumed that the pin distribution is sparse~\cite{Ong} ($R_0\gg a_B$, where 
$R_0$ is the average pin separation, and $a_B = \sqrt{\phi_0/B}$ the lattice spacing).  Appendix C 
summarizes the solution of OW.

In response to the driving current ${\bf J}e^{j\omega t}$, the magnitude and phase of the vortex velocity 
(averaged over the sample) produces the ac voltage that determines $\hat{\rho}$.  At low $\omega$, the 
inter-vortex forces dominate the friction force $\eta \dot{\bf u}$, so the velocity response is predominantly 
inductive ($\omega{\cal L}_v \gg\rho_1$).  In the opposite limit of large $\omega$, $\eta \dot{\bf u}$ is 
dominant. As ${\cal L}_v\rightarrow 0$, $\rho_1$ approaches the free-flow value $B\phi_0/\eta$.  Thus 
${\cal L}_v$ is a measure of the average restraining force on each vortex.  It is largest if the restraining 
forces are dominant, but vanishes when viscous damping dominates.  

The main feature of the solution is that the lattice propagator $G({\bf R},\omega)$, which transmits 
information on the vortex displacement at site $\bf R$ to its surroundings, has an {\em effective range} 
$R_G = a_B\sqrt{c_{66}/\eta\omega}$ that varies as $1/\sqrt{\omega}$ (Eq. \ref{p}).  At low 
frequencies, $R_G\gg a_B$, which implies that the motion of any one vortex is strongly correlated with that 
of a huge number of neighbors, a subset of which are pinned (the 2D correlation volume is $\sim R_G^2$).  
Hence the velocity response is large and inductive.  With increasing $\omega$, the average vortex is 
restrained by fewer and fewer pinned vortices as $R_G$ shrinks.  The in-phase (dissipative) component of 
the average velocity increases, while the inductance drops.  When $\omega$ exceeds the characteristic 
frequency 
\begin{equation}
\omega_{66}= \frac{4\pi c_{66}}{\eta}\frac{a_B^2}{R_0^2},
\label{w66}
\end{equation}
$R_G$ falls below the average pin spacing $R_0$.  The majority of vortices now respond as if they are 
free, and $\rho_1$ rapidly approaches the Bardeen-Stephen free flux-flow value $\rho_f = B\phi_0/\eta$ 
while ${\cal L}_v$ decreases to zero.  

The crossover is observed as the knee in  $\rho_1$ at $T$ just below $T_m$ (see Fig. \ref{rho1}a).  As 
$T$ approaches $T_m$ from below, the rapid softening of $c_{66}$ causes this crossover to occur at 
progressively lower frequencies.  Finally, in the liquid state above $T_m$, the vanishing of $c_{66}$ implies 
that virtually all the vortices are decoupled from the pins: $\rho_1$ equals $\rho_f$ at all $\omega$ in the RF 
range, while ${\cal L}_v$ is zero.

\subsection{The fitting procedure}  
As discussed in Appendix C, the OW solution has 4 adjustable parameters at each $T$, viz. $\kappa(T)$, 
$c_{66}(T)$, $\eta(T)$ and $R_0$.  Empirically, the fits are not sensitive to the particular choice of $R_0$ 
(as long as $R_0\gg a_B$).  Following WOGT, we set $R_0 = 7\;a_B$ for $B$ = 1 T.

As discussed in Sec. \ref{fixedfreq}, the viscosity $\eta(T)$ may be obtained from the initial slope of 
$\rho_1$ vs. $H$ measured at 15.4 MHz (Fig. \ref{rhoeta}).  Adopting the viscosity data in the insert of 
Fig. \ref{rhoeta}, we then have only {\em two} adjustable parameters at each $T$, viz. $\kappa(T)$ and 
$c_{66}(T)$ [equivalently, $\omega_p(T)$ and $\omega_{66}(T)$].  As we have a {\em continuous} 
spectrum measured over 2 decades in frequency to find 2 numbers, the fitting problem is evidently strongly 
over-constrained.  

The fitting procedure consists in selecting seed values for $\omega_p$ and $\omega_{66}$ so that the 
quantity $\hat{\cal{S}}(\omega)$  (Eq. \ref{S}) can be computed by summing the propagator 
$G_{2D}({\bf R_i},\omega)$ (Eq. \ref{G2Da}) over an area much larger than $R_G^2$ (the pins are 
arrayed in a triagular superlattice of spacing $R_0$, and the sum is typically extended to 12 impurity `shells' 
from the origin).  With $\hat{\cal{S}}(\omega)$, we may compute the complex resistivity 
$\hat{\rho}(\omega)$, viz. 
\begin{equation}
{\hat\rho}(\omega)  =  \left( \frac{B\phi_0}{\eta}\right)
\left[ 1 - \left( \frac{\omega_p a^2 } {{\rm j}\omega R_0^2}  \right) 
\frac{1}{1+\hat{\cal{S}}(\omega)}\right].
					\label{rho}
\end{equation}

The calculated curve is then compared with the measured spectra of $\rho_1(\omega)$, and the fit 
parameters are changed in an iterative manner until the fit is optimized (the inductance data are not used in 
the optimization).  For spectra near $T_m$, the high-frequency curvature is most sensitive to $\omega_p$ 
whereas the curvature at low $\omega$ is largely determined by the choice of $\omega_{66}$, as expected 
on physical grounds (the value of $\eta$ mainly fixes the overall scale of $\rho_1$).  As shown in Fig. 
\ref{rho1}b, close fits to the data are achieved over the 2-decade range of $\omega$.  At low $\omega$, 
the calculated $\rho_1$ is capable of matching the curvature of the measurements to a surprising degree 
(with only 2 adjustable parameters).  The close agreement achieved in Fig. \ref{rho1}b offers encouraging 
support for the validity of the model.

\subsection{Calculated inductance}
\begin{figure}
\includegraphics[width=7cm]{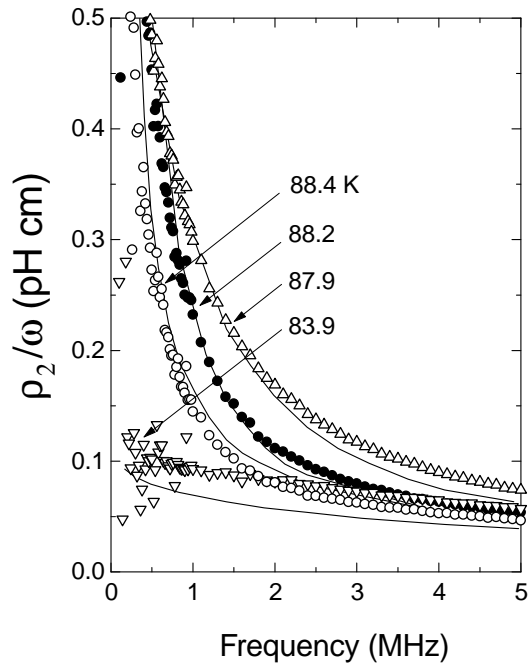}
\caption{\label{FitL} Comparison of calculated $\rho_2(\omega)/\omega$ (solid lines) with the measured 
vortex inductance ${\cal L}_v(\omega,T,H_0)$ (open symbols) over the frequency range 100 kHz to 20 
MHz in a field $H_0$ = 2 T.  $\rho_2(\omega)/\omega$ is calculated from Eq. \ref{rho} using the values of 
$\omega_p$ and $\omega_{66}$ derived from the fits to $\rho_1$ in Fig. \ref{rho1}b.  The estimated 
volume inductance has been subtracted from the total sample inductance (see text).  Above 87 K, ${\cal 
L}_v$ strongly diverges as $\omega\rightarrow 0$, while at 83.9 K, the divergence is not 
observed.}

\end{figure}

With the parameters $\omega_p$ and $\omega_{66}$ optimized at each $T$, we may calculate the 
vortex-lattice inductance ${\cal L}_v$.  As discussed above, in the limit $\omega\rightarrow 0$, the 
extended range of the propagator $G_{2D}(R,\omega)$ implies that each vortex is restrained by a great 
many pins.  Hence the velocity response is purely inductive.  The calculated ${\cal L}_v$ displays the strong 
divergence observed in the RF measurements.  In Fig. \ref{FitL}, we compare at 4 selected $T$ the 
calculated ${\cal L}_v$ (solid lines) with the observed ${\cal L}_v$ (open symbols).  

The sample inductance is the sum of the vortex-lattice term ${\cal L}_v$ and the volume inductance ${\cal 
L}_{vol}$, viz.
\begin{equation}
{\cal L}_s(\omega,T,H) = {\cal L}_v(\omega,T,H) + {\cal L}_{vol}(\omega,T,H).
\label{Ls}
\end{equation}
The volume term cannot be measured directly at low $\omega$ because it is much weaker than the vortex 
term.  However, at frequencies above $\sim$ 3-10 MHz (depending on $T$), it accounts for nearly all of 
the sample inductance (see Sec. \ref{fixedfreq}).  Because of this uncertainty, we used the spectra of 
$\rho_1$ to optimize the fit parameters.

The contribution of the volume inductance has to be estimated to isolate the vortex-lattice term (Eq. 
\ref{Ls}).  In Sec. \ref{fixedfreq}, this was carried out using the observed $\rho_1$ together with 
calculations for a sample with elliptical section.  The estimated ${\cal L}_{vol}$ has been subtracted to 
isolate the vortex term ${\cal L}_v$ in the plots displayed in Fig. \ref{FitL}.  Although the uncertainties in 
the estimated ${\cal L}_{vol}$ are largest at low $\omega$, they have the least impact on the comparison 
because ${\cal L}_s$ diverges steeply in this limit.  The comparison (inset) shows good agreement between 
calculation and measurement above 87 K (no further refinements of the fit parameters were made in 
comparing with ${\cal L}_v$).  As $T$ approaches $T_m\simeq$ 88.5 K, the low-$\omega$ divergence in 
the observed ${\cal L}_v$ progresses to lower frequencies (see the curves at 87.9, 88.2 and 88.4 K in 
inset).  The calculated curves also match this progression.  At low $T$, however (83.9 K curve), the 
agreement is not as good because the uniform-$J$ assumption used to extract $\hat{\rho}$ is increasingly 
suspect.

\section{\label{modulus} Fit results}
\subsection{Collapse of the shear modulus}
\begin{figure}   
\includegraphics[width=9cm]{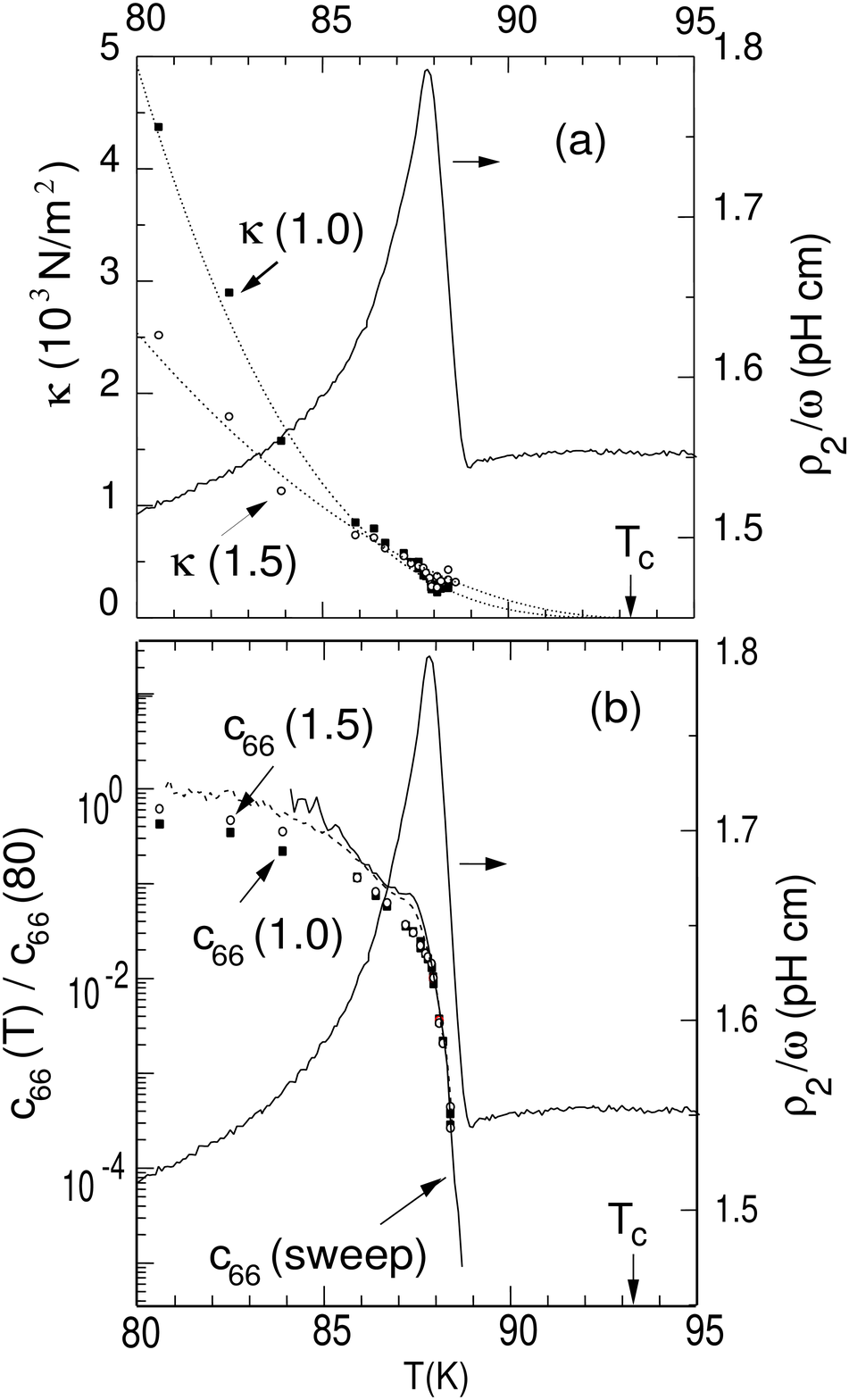}
\caption{\label{figc66} (a) Values of the Labusch parameter $\kappa$ obtained from the fits, using either 
$\eta_a$ (open circles) or $\eta_b$ (solid squares). The sharp peak (right scale) is the behavior of the 
sample inductance ${\cal L}_s = \rho_2/\omega$ measured at 1 MHz in a 2-T field [the zero of ${\cal 
L}_s$ is arbitrary (Sec. \ref{fixedfreq})]. (b) The normalized shear modulus $c_{66}$ (log scale) versus 
$T$ obtained from the fits, using either $\eta_a$ (open circles) or $\eta_b$ (solid squares) in Fig. 
\ref{rhoeta}.  Near $T_m$, $c_{66}$ decreases by 4 orders of magnitude within an interval of 0.5 K.  The 
solid line represent values of $c_{66}$ obtained from fits to $\hat{\rho}(\omega)$ with $\omega$ fixed at 1 
MHz, and $H$=2 T, using $\eta_b$.  The dashed line is obtained with $\omega$ fixed at 2 MHz, but using 
$\eta_a$.
}
\end{figure}

The values of $\kappa(T)$ and $c_{66}(T)$, obtained from $\omega_p(T)$ and $\omega_{66}(T)$, 
respectively, are shown in panels (a) and (b) of Fig. \ref{figc66}.  In each case, we show two sets of values 
depending on whether $\eta_a$ or $\eta_b$ has been used in the fits.  We interpret the decrease of 
$\kappa$ as $T\rightarrow T_c$ (Panel (a)) as reflecting the decrease in condensation energy.  We note 
that $\kappa$ remains finite at the melting temperature $T_m$.  Hence our fit strongly argues against models 
in which the transition at $T_m$ is interpreted in terms of a vanishing $\kappa$.  

The critical parameter is the shear modulus $c_{66}$ which undergoes a steep collapse by more than 3 
decades within an interval of 1 K.  The collapse occurs over the background (gradual) decrease of 
$c_{66}$ associated with the power-law decrease of the condensate density as $T\rightarrow T_c$.  
Because of the rapid collapse in $c_{66}$, the steep increase in ${\cal L}_s$ at temperatures below 
$T_m$ is interrupted.  Within a 1-K interval, ${\cal L}_s$ drops to its normal-state value (background 
curve in Fig. \ref{figc66}b).

Each of the discrete data points for $c_{66}$ is derived from a full spectrum for $\rho_1$.  To follow the 
collapse in more detail, we have also adopted a different procedure, using the 1 MHz curve for ${\cal 
L}_s$ and $\rho_1$ measured {\em continuously} vs. $T$, together with interpolated values of $\eta_a$ 
and $\eta_b$.  This allows us to determine $c_{66}$ as a continuous curve very close to $T_m$ (the fitting 
procedure becomes unstable at lower $T$).  We show these continuous curves as solid and broken lines in 
Fig. \ref{figc66}b.  At low $T$, the continuous-fit values are slightly larger than the spectra-based fits which 
are more accurate, but close to $T_m$, they match very well.  The continuous curve allows us to monitor 
the decrease of $c_{66}$ over 4 decades.

From Fig. \ref{figc66}, we conclude that the transition at $T_m$ is associated with a rapid collapse of the 
shear modulus.  However, the Labusch parameter $\kappa$ remains finite.  This provides firm evidence that 
the transition involves the {\em collapse of the shear modulus}, rather than the vanishing of the Labusch 
parameter. 

\subsection{Summary}
In summary, we have performed measurements of $\hat{\rho}(\omega)$ of the vortex lattice in an untwinned 
YBCO crystal to investigate the collapse of the lattice inductance at the melting temperature, using a 
high-resolution 4-probe RF technique.  At all $T$ investigated, the resistivity $\rho_1(\omega)$ is strongly 
dispersive.  Over the frequency range 100 kHz - 20 MHz, $\rho_1$ increases from zero to the free-flow 
value $B\phi_0/\eta$, with a cross-over frequency scale ($\omega_{66}$) that rapidly decreases towards 
zero as $T$ approaches the melting temperature $T_m$ from below.  In the solid phase, the inductance 
$\rho_2/\omega$ displays a steep divergence as $\omega\rightarrow 0$.  However, at $T_m$, this 
low-frequency divergence collapses to give an $\omega$-independent inductance in the vortex-liquid state 
above $T_m$.  The observed spectra are qualitatively different from those in thin-film YBCO, and 
incompatible with the predictions of vortex glass theory.  To extract the vortex-lattice shear modulus from 
these spectra, we have used the mean-field solution of the Schmid-Hauger-Larkin-Ovchinikov model to fit 
the complex resistivity spectra.  From the 2-parameter fit at each temperature, the normalized shear modulus 
$c_{66}(T)$ displays a remarkable 4-decade collapse towards zero at $T_m$.  Hence the observed 
collapse of the inductance is driven by the vanishing of the shear modulus rather than the vanishing of the 
pinning strength $\kappa$.

\section*{\label{expt} Appendix A: Experimental details}
We describe first the 2-probe method used by WOGT \cite{Wu2}.  The sample $Z_2$ bridges the inner 
and outer conductors of both the incident and transmission coax cables of line impedance $Z_0$ (Fig. 
\ref{expt}a).  The load resistors $Z_1$ and $Z_1'$ are 100-$\Omega$ thin-film resistors.  This 
configuration maximizes the sensitivity of the transmitted wave to small changes in $Z_2$ when $|Z_2|\ll 
Z_0$.  The incident wave is transmitted with transmission coefficient $\Gamma_T$, and reflected with 
reflection coefficient $\Gamma_R$, where
\begin{eqnarray}
\Gamma_T & = &\frac{2Z_0Z_2}{Z_2(2Z_0+Z_1+Z_1')+(Z_0+Z_1)(Z_0+Z_1')},\\
\Gamma_R & = & \Gamma_T \frac{Z_0+Z_1'}{Z_0+Z_1}+\frac{Z_1-Z_0}{Z_1+Z_0}.
\label{gamma2}
\end{eqnarray}
Phase-sensitive detection of the transmitted signal allows $Z_2$ to be determined (Fig. \ref{expt}a).  The 
incident power from a synthesizer (Hewlett Packard 3336C) is typically fixed at +4.5 dBm, corresponding 
to a current less than 5 mA at the sample. The transmitted signal is phase-detected by a high-frequency 
lock-in Extender/Enhancer (Palo Alto Research PAR100), which down-converts signals in the range 1-25 
MHz to 1-50 kHz, suitable for a standard lock-in amplifier (Stanford Research SR830). 

\begin{figure}
\includegraphics[width=7cm]{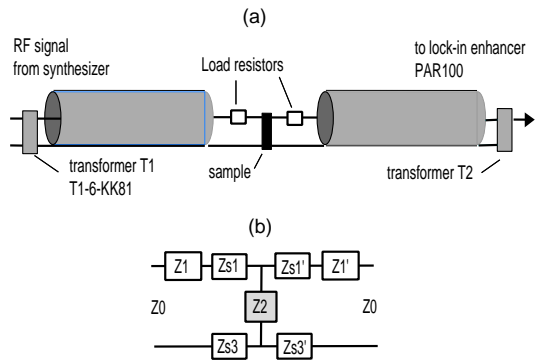}
\caption{\label{expt} (a) Schematic diagram of the measurement circuit.  The sample shunts the inner and 
outer conductors, so that the transmitted signal is highly sensitive to slight changes in the sample impedance 
$Z_2$. Transformers T1 and T2 are inserted to decouple grounds in the 4-probe technique.  The lower 
figure (b) shows the labelling of the contact impedances $Z_{si},Z_{si}'$, which may greatly exceed $Z_2$ 
in amplitude.}
\end{figure}

To modify the technique for 4-probe measurements (Fig. \ref{expt}a), we inserted an RF transformer T1 of 
bandwidth 10 kHz - 150 MHz (Mini-Circuits T1-6-KK81) at the output of the synthesizer.  The transmitted 
signal passes through a second transformer T2 before entering the PAR100.  In Fig. \ref{expt}b, the contact 
impedances are represented by the impedances $Z_{si}$ and $Z_{si}'$. The transmission and reflection 
coefficients are now
\begin{eqnarray}
\Gamma_T & = &\frac{2\xi_2}{(1+\xi_T)(1+\xi_T')-\xi_2^2},\label{gamma4t}\\
\Gamma_R & = &\frac{(\xi_T-1)(1+\xi_T')-\xi_2^2}{(1+\xi_T)(1+\xi_T')-\xi_2^2},
\label{gamma4r}
\end{eqnarray}
where $\xi_T = (Z_1+Z_{s1}+Z_2+Z_{s3})/Z_0, \;\;\xi_T' = (Z_1' + Z_{s1}'+Z_2+Z_{s3}')/Z_0 $, and 
$\xi_2=Z_2/Z_0$.

While the 4-probe method provides much higher resolution, a drawback is that the inserted transformers 
introduce strong reflections.  The largest reflection comes from the upstream transformer T1.  A wave 
reflected from the sample is reflected again at T1 and adds an $\omega$-dependent contribution to the 
original transmitted wave.  

The effect of the multiply reflected wave on the total transmitted signal at the detector is 
expressed as 
\begin{equation}
\Gamma_T^{obs} \equiv \frac{V_{out}}{V_{in}} = {\rm e}^{jk\ell_{tot}}\Gamma_T\Gamma_T^1
\left[1 + {\rm e}^{j2k\ell_1} \Gamma_R\Gamma_R^1 + \cdots \right],
\label{mult} 
\end{equation}
where $\Gamma_T^1$ and $\Gamma_R^1$ are, respectively, the transmission and reflection coefficients of 
T1.  The path-lengths from T1 to the detector, and from T1 to the sample, are called $\ell_{tot}$ and 
$\ell_1$, respectively.  Higher-order reflection contributions (notably from T2) are indicated by $\cdots$.  In 
addition, we are now also sensitive to slight $\omega$-dependent deviations from unity of the PAR100's 
transfer function $\hat{g}(\omega)$. 

To compensate for these two background contributions, we used a reference combination 
$Z_r,Z_{1,r},Z'_{1,r}$ tailored to have a reflection coefficient $\Gamma_R^r$ nearly equal to that of the 
sample (hereafter, $\Gamma_R^s$).  As $ |Z_s| \ll 1$, the sample's $\Gamma_R^s$ is very close to 0.33, 
with $Z_1$ and $Z_1'$ = 100 $\Omega$. The best choice for the reference combination is $Z_r$ = 10 
$\Omega$, and $Z_{1,r}=Z'_{1,r}$ = 90.9 $\Omega$, which has an $\omega$-independent 
$\Gamma_R^r$ of 0.3344, and a transmission coefficient $\Gamma_T^r$= 0.044.  

With the sample in place, the signal presented at the input of the SR 830 lock-in is given by $V_s = 
\hat{g}(\omega)V_{out}(Z_s)$, where $V_{out}$ is given by Eq. \ref{mult}.  If the sample (and load 
resistors) are replaced by the reference combination, we have $V_r = \hat{g}(\omega)V_{out}(Z_r)$.  
Dividing these two equations removes $\hat{g}(\omega)$.  This leaves the multiple reflection factor (quantity 
in $[\cdots]$ in Eq. \ref{mult}).  However, as $\Gamma_R^s$ and $\Gamma_R^r$ are nearly identical by 
design, the leading term in $[\cdots]$ is the same (the other terms are down by a factor $\Gamma_T^r\ll 
1$).  Hence we have 
\begin{equation}
\frac{V_s}{V_r} = \frac{\Gamma_T^s}{\Gamma_T^r}[1 + {\cal O}(\Gamma_T^s)],
\label{vs}
\end{equation}
where $\Gamma_T^s$ and $\Gamma_T^r$ are given by Eq. \ref{gamma4t} with $Z_2=Z_s$ and $Z_r$, 
respectively.  Our procedure is to measure and store the curve of $V_r$ vs. $\omega$ at each temperature 
of interest.  Then $V_s$ is measured versus $\omega$.  Since $\Gamma_T^r$ is known, $\Gamma_T^s$ 
(and $Z_s$) may be calculated from $V_s$ and $V_r$ using Eq. \ref{vs}.  We checked the reliability of the 
procedure using a second reference consisting of just a single 50-$\Omega$ resistor connecting the inner 
conductors of the two coax cables (this also has $\Gamma_R^r$ = 0.333, but a much larger 
$\Gamma_T^r$ = 0.667). 

\section*{\label{impedance} Appendix B: Impedance of a conductor with elliptical 
cross-section}
The impedance $Z_s(\omega)$ of a conductor with elliptical cross-section is obtained by solving the 2D 
Helmholtz equation $(\nabla^2+\hat{\kappa}^2){\bf A}=0$ in elliptical coordinates $(\xi,\eta)$, where 
$\hat{\kappa}^2 = 2j\delta_n^{-2}$.  With ${\bf A}(\xi,\eta) = \psi(\xi)\phi(\eta)\hat{\bf z}$, the Helmholtz 
equation separates into the two Mathieu equations, viz. \cite{McL}
\begin{eqnarray}
\frac{d^2\psi}{d\xi^2} - [a(\hat{q}) -2\hat{q}\cosh 2\xi]\psi = 0,\\
\frac{d^2\phi}{d\eta^2} + [a(\hat{q}) -2\hat{q}\cos 2\eta]\phi = 0,
\label{Mathieu}
\end{eqnarray}
where $\hat{q}\equiv -\hat{\kappa}^2h^2/4$ ($h$ is the foci spacing).  With the boundary condition 
$A(\xi_0,\eta) = A_S$ (i.e. a constant on the surface $S$), the vector potential may be expanded in terms 
of the Matthieu functions as
\begin{equation}
A(\xi,\eta) = \sum_{n=0}^{\infty} c_{2n}Ce_{2n}(\xi,\hat{q})ce_{2n}(\eta,\hat{q}).
\end{equation}
$Ce_{2n}$ is expressed by the expansion 
$Ce_{2n}(\xi,\hat{q})= \sum_{r=0}^\infty A^{(2n)}_{2r} \cosh(2r\xi)$
[the same expansion holds for $ce_{2n}$, but with $\cos(2r\eta)$ in place of $\cosh(2r\xi)$].  

The Mathieu functions are not tabulated because the separation parameter $a(\hat{q})$ and the coefficients 
$A^{(2n)}_{2r}$ are functions of $\hat{q}$.  For a given $\hat{q}$, $a(\hat{q})$ satisfies the continued 
fraction  
$a = -(\hat{q}^2/2)/[1-(a/4)-(\hat{q}^2/64)/[1-\cdots)]]$ \cite{McL}.  We truncated the continued 
fraction at the 12$^{th}$ order.  The 12 roots \{$a_n(\hat{q})$\} generate the full matrix 
$A^{(2n)}_{2r}$ with $(n,r)\;=1,\cdots 12$.  

Finally, with the ac voltage $V(\omega) = -j\omega A_S$ and the ac current $I = 
-\hat{\kappa}^2/\mu_0\int_S d\xi d\eta A(\xi,\eta)$, the impedance may be expressed as 
\begin{equation}
Z_s(\omega) = \frac{j\omega\ell\mu_0}{(2\pi)^2} \left[ \sum_n \frac{(A^{(2n)}_0)^2}{{\cal L}_{2n}} 
\frac{Ce_{2n}'(\xi_0,\hat{q})} {Ce_{2n}(\xi_0,\hat{q})} \right]^{-1},
\end{equation}
where $Ce_{2n}'= \partial Ce_{2n}/\partial \xi$, and ${\cal L}_{2n} = \int_0^{2\pi} 
ce^2_{2n}(\eta,\hat{q})d\eta$.  In Fig. \ref{elliptical}, we display the $\omega$ dependence of 
$R_s(\omega)$ and $\cal{L}(\omega)$ calculated with parameters $\xi_0\simeq 0.108$ and $h= 250 
\mu$m, matching the sample cross-section, and $\rho_1 = 4{\rm \mu\Omega cm}$.

\section*{\label{Ong} Appendix C: The solution of Ong and Wu}
We summarize the equations in Ref. \cite{Ong}.  The displacement ${\bf u_l}$ of a vortex at site $\bf l$ is 
described by the equation
\begin{equation}
\eta \dot{\bf u_l} + 
\sum_{s,\bf m} {\bf D}^s_{\bf{l,m}} \cdot {\bf u_m}
+ \kappa\sum_{pins\;\bf i} {\bf u_i} \delta_{{\bf l,i}} 
= {\bf J \times \hat{z}}\phi_0.
		\label{eta}
\end{equation}
Only a subset of vortices (at sites indexed by $\bf i$) are affected by the pins.  The remaining vortices are 
`free', apart from being restrained by the lattice forces represented by ${\bf D}^s_{\bf{l,m}}$ ($s= T, L$ 
indexes the polarization mode).  The displacement at an arbitrary site $\bf l$ is related to the displacement of 
the pinned vortices by
\begin{eqnarray}
{\bf u_l}(\omega) & = & \left( \frac{J\phi_0}
{\rm j\omega\eta}\right){\bf\hat y} - \frac{\kappa}{N} \sum_{pins\;\bf i} 
{\bf u_i}(\omega)\cdot  \nonumber\\
&  &  \times\sum_{s,{\bf q}}
\frac { {\bf\hat e}_s({\bf q}) {\bf \hat e}_s({\bf q}) 
{\rm e}^{ j\bf q \cdot({\bf R_l}- {\bf R_i} ) } }
{ [D_s(\bf q) + \rm j\omega\eta] },
					\label{ul}
\end{eqnarray}
where ${\bf {\rm\hat{e}}}_s({\bf q})$ is the unit vector for the mode $s$.  

At the mean-field level, the displacement of the vortex at a pin-site ${\bf u}_{imp}$ is
\begin{equation}
{\bf u}_{imp}(\omega)  = {\bf\hat y} \left( {J\phi_0\over{\rm j\omega\eta}} 
\right) \frac{1}{1+ \hat{\cal{S}}(\omega)},
\label{uimp}
\end{equation}
where all the interaction effects are in the term
\begin{equation}
\hat{\cal{S}}(\omega) \equiv \frac{\omega_p a^2 }{{\rm j}\omega R_0^2} + 
\kappa[ G(0,\omega) + \sum_{\bf i}' G({\bf R_i},\omega)],	
			\label{S}
\end{equation}
and $\omega_p \equiv \kappa/\eta$ is the pinning frequency.  The lattice propagator  $G({\bf R},\omega)$ 
is defined as \cite{Schmid,Ong}
\begin{equation}
G({\bf R},\omega) = \frac{1}{N} \sum_{s{\bf q}} 
\frac { ({\bf{\hat y} \cdot {\hat e}}_s({\bf q}) )^2
{\rm e}^{ {\rm j}{\bf q\cdot R} } } 
{ [D_s({\bf q}) + {\rm j}\omega\eta]}.
					\label{eq:G}
\end{equation}
In Eq. \ref{S}, the first term represents the uniform $\bf q$=0 mode of the transmitted elastic forces.  The 
remaining terms are elastic forces transmitted to the pinned vortex at the origin $\bf 0$ arising from its own 
displacement ($G(0,\omega)$), or the displacement of other pinned vortices $\sum_{\bf i}' G({\bf 
R_i},\omega)$ (the prime indicates that the $\bf i=0$ term is left out).

Neglecting the longitudinal modes, the transverse force matrix is expressed as \cite{Schmid}
\begin{equation}
D_T({\bf q}) = [c_{66}  q^2 + c_{44} q_{z}^2]a_B^2.  \label{DT}
\end{equation}
where $q$ is the in-plane component of ${\bf q}$, viz. ${\bf q} = (q,q_z)$, and $c_{66}$ and $c_{44}$ 
are the shear and tilt moduli, respectively.  

In the 2D limit, the propagator simplifies to the form (after angular-averaging within the $ab$ plane and 
neglecting the tilt modulus $c_{44}$) 
\begin{equation}
\kappa G_{2D}({\bf R}_{\perp},\omega) = g \int_{0}^{Q}dq 
\frac{qJ_0(qR) } { [q^2 + {\rm j} p^2] }.
				\label{G2D}
\end{equation}
\noindent
where $Q=\sqrt{4\pi/a_B^2}$, $J_0(x)$ is the zeroth order Bessel function, $R=|{\bf R}_{\perp}|$, with 
${\bf R = (R}_{\perp}, R_z)$, and $g$ the coupling constant $\kappa/4\pi c_{66}$. The frequency 
dependence appears only in the characteristic wavevector 
\begin{equation}
p = \sqrt{ {\eta\omega\over{c_{66}a_B^2} } }   \label{p} 
\end{equation}
which serves as a cut-off of the logarithmic divergence in the 2D case.  Setting $R=0$ in Eq. \ref{G2D}, we 
have 
\begin{equation}
\kappa G_{2D}(0,\omega) = g \left[ \ln \sqrt{\left( 1+ 
{Q^4\over{p^4}} \right)}  - {\rm j} \arctan{({Q^2\over{p^2}})} \right].    
			\label{G2D0}
\end{equation}
As the propagator $G_{2D}(R,\omega)$ is used repeatedly in the sum in Eq. \ref{S}, it is convenient to 
adopt the approximation (accurate to a few percent)  
\begin{eqnarray}
\kappa G_{2D}^a(R,\omega) &=& g  \left\{ \ln \sqrt{\left[\frac
{(\gamma R)^4 + p^{-4}} {(\gamma R)^4 + Q^{-4} } \right]} \right. \nonumber\\
&& \left. -{\rm j} \arctan({Q^2\over{p^2}}) \frac{1}
{ [1+(\beta pR)^2]^2 } \right\},
 		\label{G2Da}
\end{eqnarray}
\noindent
with $\beta = 0.52$ and $\gamma = 0.85$.  Equations \ref{G2D0} and \ref{G2Da} were given incorrectly 
in OW \cite{Ong} ($cpf$ their Eqs. 18 and 19).

\section*{Acknowledgments}
We thank Hui Wu for generous assistance and helpful advice.  N.P.O. acknowledges support from the U.S. 
Office of Naval Research (Contract N00014-01-0281) and the New Energy and Industrial Tech. Develop. 
Org. (NEDO), Japan.  R.G. and L.T. are funded by NSERC of Canada and FCAR of Quebec. L. T. 
acknowledges support from the Canadian Inst. for Advanced Research and the A.P. Sloan 
Foundation.  

\bigskip\noindent
$^*${\em Present address of P.M.}: Koch Industries, 20 East Greenway Plaza, Houston, TX 
77046.\newline
$^\dagger${\em Present address of L. T.: Department of Physics,
University of Toronto, Toronto, Ontario, CANADA  M5S 1A7.}

\end{document}